\documentclass[10pt,journal,compsoc]{IEEEtran}
\usepackage{amsmath,amssymb,amsfonts}
\usepackage{array}
\usepackage[table]{xcolor}
\usepackage{graphicx}
\usepackage{svg}
\usepackage{booktabs}
\usepackage{longtable}
\usepackage{colortbl}
\usepackage{multirow}
\usepackage{csquotes}
\usepackage{xspace}
\usepackage{comment}
\usepackage{url}
\usepackage[numbers]{natbib}
\usepackage{hyperref}
\usepackage{balance}
\usepackage{float}
\usepackage{capt-of}
\graphicspath{{./}{SoundFleXR_Study/}}

\definecolor{contribpurple}{HTML}{E9E1F8}

\newcommand{\condFull}{\textsc{Full HRTF}\xspace}
\newcommand{\condAzim}{\textsc{Azimuth-Only HRTF}\xspace}
\newcommand{\condStereo}{\textsc{Two-Channel}\xspace}

\begin{document}

\title{%
  \fontsize{20}{23}\selectfont
  SoundFleXR: Selectively Constraining Axes of Audio Rendering Presentations\par
}
\author{Brandon Woodard,
        Hyunju Kim,
        Hyunsung Cho,
        David Lindlbauer,
        Rhonda Wilson,
        and Joseph LaViola Jr.%
\IEEEcompsocitemizethanks{%
  \IEEEcompsocthanksitem
  Brandon Woodard is with Brown University.
  \IEEEcompsocthanksitem
  Hyunju Kim is with Cornell University.
  \IEEEcompsocthanksitem
  Hyunsung Cho \& David Lindlbauer are with
  Carnegie Mellon University.
  \IEEEcompsocthanksitem
  Rhonda Wilson is with Dolby Laboratories.
  \IEEEcompsocthanksitem
  Joseph J. LaViola Jr. is with the University of Central Florida.
}}

\newsavebox{\teaserbox}
\sbox{\teaserbox}{
  \begin{minipage}{0.9\textwidth}
    \centering
    \includegraphics[width=\linewidth]
      {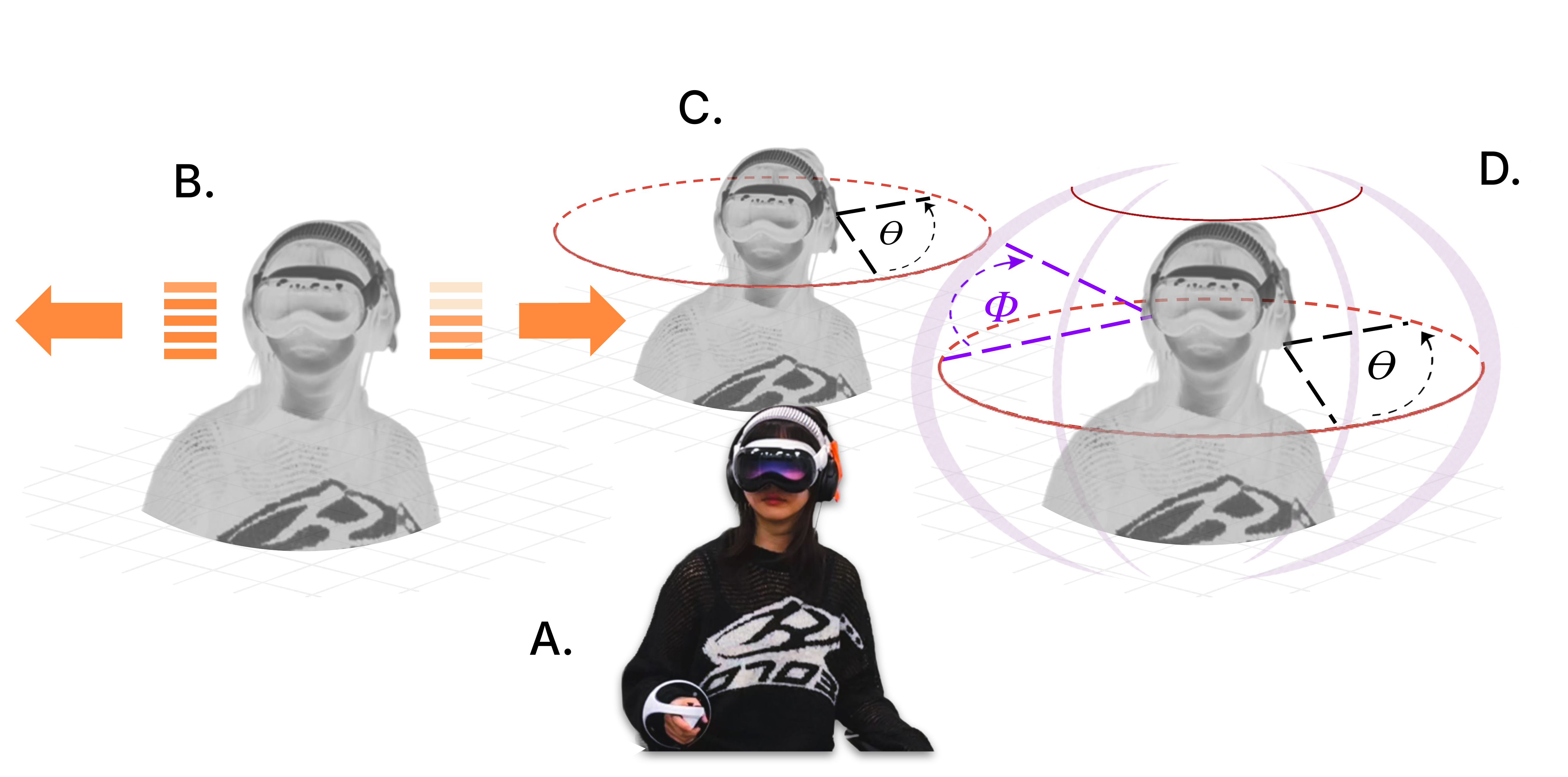}

    \captionof{figure}{
      \textbf{SoundFleXR:} Comparison of audio rendering conditions across
      XR localization tasks with varying degrees of freedom.
      (A) Hardware: Apple Vision, Bose QC Headphones, and PSVR2 controllers
      (B) Stereo/2-Channel Localization
      (C) HRTF with Azimuth ($\theta$) cues
      (D) HRTF with both Azimuth and Elevation ($\phi$) cues
    }
    \label{fig:teaser}
  \end{minipage}
}

\IEEEtitleabstractindextext{
  \noindent\usebox{\teaserbox}
  \vspace{0.5\baselineskip}

  \begin{abstract}
        Extended reality (XR) systems increasingly rely on audio cues to help users locate, interpret, and interact with virtual content, especially when the content is outside the field of view. To support this, many systems use head-related transfer functions (HRTFs) to approximate real-world acoustics. However, a realistic simulation is only one possible presentation. Audio can also be altered by constraining, flexing, or contracting the spatial dimensions represented in the signal. This work examines how selectively constraining or flexing the number of audio dimensions affects localization performance across XR task structures with different spatial degrees of freedom. We compare two-channel spatial audio, azimuth-only HRTF, and full HRTF rendering across orientation-based (3DoF) and translation-based (6DoF) tasks. Azimuth-only HRTF was mutually non-inferior to full HRTF for total angular error across task conditions, including at the strictest tested tolerance in 3DoF ($5^\circ$), and remained comparable for azimuth error at ($10^\circ$) and larger tolerances in both 3DoF and 6DoF. In contrast, two-channel stereo required coarser tolerances to reach non-inferiority, becoming comparable to full HRTF for 6DoF azimuth error at ($15^\circ$) but not ($10^\circ$). From these comparisons, we derive design lessons and application scenarios that indicate when fewer spatial cues provide sufficient support and when richer spatial cues are beneficial.
  \end{abstract}

  \begin{IEEEkeywords}
    spatial audio, extended reality, audio localization, HRTF
  \end{IEEEkeywords}
}
\IEEEaftertitletext{\vspace{-16pt}}
\maketitle
\IEEEdisplaynontitleabstractindextext
%\section{Introduction}
%\label{sec:intro}
\IEEEraisesectionheading{\section{Introduction}\label{sec:intro}}
The use of audio cues is an important method for helping users locate events, move or turn towards them, and stay aware of what is happening around them throughout a variety of XR experiences. Such applications often use head-related transfer functions (HRTFs), which approximate real-world acoustics by encoding direction-dependent spatial cues~\cite{wightman1989headphone,wenzel1993localization,begault2001direct}. Prior work demonstrates spatial audio as a method to improve a user's ability to search, navigate through a space, and immerse themselves in an experience compared to non-spatialized audio~\cite{ruminski2015experimental, poeschl2013integration, di2025audiovisual} and has become a standard practice in producing realistic experiences in XR.

 In particular, HRTFs model the direction-dependent filtering imposed by the listener's head, torso, and pinnae, enabling binaural rendering that can remain listener-relative and support 6DoF interaction in XR~\cite{wightman1989headphone,begault2001direct}. Despite this expanded design space in XR, there is little guidance on how audio should be presented for different interaction goals. Typically, higher-fidelity or more realistic rendering is often treated as the default target, implicitly positioning realism as the primary measure of a good auditory interface.

The \condFull condition used in this study encodes azimuth, elevation, spectral shaping, and distance---but human sensitivity to these dimensions is not uniform. Horizontal localization relies on binaural cues, specifically interaural time and level differences, and is well understood and reliable~\cite{wightman1989headphone}. Elevation localization, by contrast, relies on weaker monaural spectral cues that are particularly susceptible to error when non-individualized HRTFs are used~\cite{wenzel1993localization,romigh2014isolating,jenny2021trust}. This asymmetry suggests that rendering formats which preserve azimuthal structure while contracting other dimensions may remain non-inferior, or not meaningfully worse within a chosen threshold, to \condFull for tasks whose spatial demands are primarily horizontal.

A further consideration is that XR tasks vary substantially in their spatial DoF. Orientation-based tasks, such as turning toward a notification or attending to a speaker in a social XR context, constrain users to rotation only (3DoF), making azimuth the primary actionable cue. Translation-based tasks, such as searching for a hidden object in a physical space, afford additional movement strategies including walking toward a source, using distance attenuation, and resolving ambiguity through viewpoint shift (6DoF). This difference suggests that the informational requirements of audio rendering may vary with task type, and that the relationship between rendering dimensionality and localization performance is unlikely to be uniform across interaction contexts.

Inspired by prior work on interaction constraints in 3D user interfaces and VR, particularly by Stuerzlinger et al., we view constraints not only as restrictions but as design strategies that can reduce unnecessary DoF and better align interaction with task demands~\cite{stuerzlinger2011value, stuerzlinger2010value, hayatpur2019planeraypoint, mendes2016benefits, smith2001integration}. Related fidelity accounts formalize \emph{input fidelity} and \emph{output fidelity} relative to a reference and treat their combination as \emph{system fidelity}~\cite{bonfert2025interaction}. Instead of measuring fidelity, SoundFleXR aims to investigate how a task's spatial dimensions and tolerance margins determine which audio presentation dimensions are necessary.

We therefore present the \emph{SoundFleXR} study, which examines how constraining and extending spatial audio dimensions affect localization performance, and how these effects vary between 3DoF and 6DoF conditions. We evaluate localization performance under three rendering conditions: \condStereo (ordinary two-channel stereo), \condAzim (azimuth-only HRTF), and \condFull (full HRTF), in both a 3DoF orientation task and a 6DoF search task. Both quantitative measures of localization accuracy and qualitative feedback are reported---we investigate the following research questions:
\begin{itemize}
    \item RQ1: Can simpler audio presentations with fewer spatial dimensions achieve localization performance comparable to HRTF-based in an XR setup?
    \item RQ2: What design lessons and XR application classifications emerge from the (mis)alignment between task objectives, DoF, and audio presentation dimensions?
\end{itemize}

\section{Related Work and Background}
\subsubsection{Spatial Audio in XR}
Spatial audio has been shown to improve user performance in XR environments meaningfully. Prior work demonstrates that spatialized audio enhances object detection, reduces response time, and improves head-orienting behavior relative to non-spatialized 2D audio~\cite{di2025audiovisual,ruminski2015experimental,chen2018investigating}.
Suarez et al. further compared measured and modeled HRTFs for real-time 3D audio processing in VR~\cite{suarez2017comparison}.
While these findings collectively support the value of spatial audio in XR, they largely treat spatial audio as a binary choice between non-spatialized and fully spatialized rendering. The intermediate design space, between ordinary 2-channel stereo and full HRTF, remains underexplored~\cite{di2025audiovisual}.
It is not yet well understood which spatial dimensions are necessary for a given task, or whether contracted rendering formats can provide sufficient support when full HRTF fidelity is not required.

\subsubsection{Human Sound Localization and Perceptual Limits}
Human sound localization is usually more reliable in the horizontal plane than in the vertical plane. Horizontal localization is driven by robust binaural cues, interaural time differences (ITD) and interaural level differences (ILD), which are stable across listeners and well-preserved by generic HRTFs~\cite{wightman1989headphone}.
Elevation localization, by contrast, relies on monaural spectral cues shaped by the pinna, which are both weaker and more idiosyncratic. When non-individualized HRTFs are used, these spectral cues may not match the listener's own pinna geometry, leading to elevated error rates and increased front-back confusion ~\cite{wenzel1993localization,romigh2014isolating,jenny2021trust}.
This asymmetry between horizontal and vertical localization motivates our decision to evaluate an azimuth-only rendering condition that preserves the more reliable horizontal cues while removing potentially misleading elevation information.

\subsubsection{3DoF and 6DoF as Ecological XR Task Classes}

XR interaction tasks can be characterized by their spatial degrees of freedom. Orientation-based tasks (3DoF) constrain users to rotation only, as in seated or simply turning toward a notification. In 3DoF, distance and translation cues are unavailable, making azimuth the primary actionable spatial cue. This suggests that ordinary 2-channel or fixed-elevation HRTF rendering is sufficient for 3DoF localization, as the additional spatial dimensions encoded by full HRTF may not contribute meaningfully to task performance when only head rotation is available.

Translation-based tasks (6DoF), by contrast, allow users to leverage motion, distance attenuation, and viewpoint shift to resolve source location. Even ordinary 2-channel audio can provide usable distance information in 6DoF via loudness attenuation~\cite{clemenson2021rethinking,woodard2025audiomixr}, and the additional embodied movement afforded by 6DoF may compensate for missing HRTF elevation cues. More broadly, head movement itself is known to support auditory disambiguation and improve localization accuracy, especially in the horizontal plane~\cite{mcanally2014headmovement,gulli2022spontaneous,rebillat2012egocentric}. This challenges the intuition that task complexity directly implies the need for richer rendering complexity.

\section{Audio Dimensions}
We define our audio presentation dimensions as the amount of directional structure the renderer preserves between the virtual source and each ear channel.

Specifically, we investigate audio presentation dimensions as a means of guidance for XR tasks. Our objective is to demonstrate how simpler or more complex audio presentations differ in performance and user experience. We detail the three conditions used in the \emph{SoundFleXR} study in Figure \ref{fig:teaser} and Table \ref{tab:active_transfer_variables}. 

The goal of the study is inspired by Stuerzlinger et al. \cite{stuerzlinger2011value, stuerzlinger2010value}'s emphasis on aligning different visual presentation dimensions in VR and task objectives instead of maximizing visual dimensions. As a first step toward understanding how audio presentation should align with XR task objectives, we structure our evaluation around angular localization, a fundamental component of spatial search, navigation, and interaction in XR~\cite{ruminski2015experimental,chen2018investigating,endsley2017augmented,woodard2025audiomixr}.

\subsection{Audio Presentations}
\label{sec:audio_conditions}

The audio presentations were selected based on established
differences between horizontal and vertical sound localization,
with azimuth judgments benefiting from robust binaural cues and
elevation judgments relying more heavily on listener-specific
spectral cues~\cite{wightman1989headphone,wenzel1993localization,
romigh2014isolating,jenny2021trust}. Building on this distinction, our
conditions progress from coarse left--right lateralization through
ordinary two-channel audio without HRTF processing, to azimuth-only
HRTF rendering, and finally to full HRTF rendering that incorporates
both azimuth and elevation cues.

We first describe the three conditions tested in the study, then
formalize their dimension counts in the transfer map
(Table~\ref{tab:active_transfer_variables}).

\subsubsection{Mathematical Representation}

For each possible listening condition, we list the relevant audio
presentation dimensions of the listener-side transfer function.
HRTFs are conventionally expressed as
\[
H_L(\omega,\theta,\phi,d),
\qquad
H_R(\omega,\theta,\phi,d),
\]
where \(\omega\) is angular frequency, \(\theta\) is source azimuth,
\(\phi\) is source elevation, and \(d\) is source distance \cite{zotkin2004renderingLocalized,brungart1999parallax}.

Table~\ref{tab:active_transfer_variables} lists the variables used
by each condition-specific transfer map to generate the final
two-channel headphone output. Distance attenuation and reverberation
were applied consistently across audio-presentation conditions and
are therefore excluded from the presentation-dimension count.

\begin{table*}[!t]
\centering
\caption{Audio-presentation dimensions for each rendering condition.
The listed dimensions correspond to the variables used by the
condition-specific transfer map. Distance attenuation and
reverberation were applied consistently across conditions and are
therefore excluded.}
\label{tab:active_transfer_variables}

\begin{tabular}{@{}
p{0.20\textwidth}
p{0.36\textwidth}
p{0.22\textwidth}
p{0.10\textwidth}
@{}}
\toprule
\textbf{Condition} &
\textbf{Formula} &
\textbf{Dimensions Counted} &
\textbf{Count} \\
\midrule

\condStereo{} &
\(Y_e(\omega) = G_e(p)X(\omega)\) &
\(\{p\}\) &
\(D = 1\) \\
\addlinespace[1em]

\condAzim{} &
\(Y_e(\omega) = H_e(\omega,\theta,\phi_0,d_0)X(\omega)\) &
\(\{\omega,\theta\}\) &
\(D = 2\) \\
\addlinespace[1em]

\condFull{} &
\(Y_e(\omega) = H_e(\omega,\theta,\phi,d_0)X(\omega)\) &
\(\{\omega,\theta,\phi\}\) &
\(D = 3\) \\

\bottomrule
\end{tabular}
\end{table*}

\subsubsection{Definition of Terms}

\begin{itemize}
    \item \(X(\omega)\) denotes the frequency-domain source signal.
    \item \(Y_e(\omega)\) denotes the rendered output for ear channel
    \(e \in \{L,R\}\).
    \item \(G_e(p)\) denotes the stereo-panning gain for ear channel \(e\).
    \item \(p\) denotes the lateral-panning dimension.
    \item \(H_e(\omega,\theta,\phi,d)\) denotes the HRTF for ear \(e\).
    \item \(\phi_0\) denotes the fixed elevation used in \condAzim.
    \item \(d_0\) denotes the nominal controlled source distance.
\end{itemize}

\subsection{Varying Degrees-of-Freedom}
To answer RQ1 and RQ2 (Section~\ref{sec:intro}), we evaluate 3DoF and 6DoF interaction, which represent two canonical motion-tracking configurations in XR~\cite{angelov2020modern, buttussi2017effects, takala2026xr}. Additionally, these setups distinguish between orientation-only and position-aware tracking, which span the reality--virtuality continuum~\cite{milgram1999taxonomy,mcmahan2018systemFidelity} and may support the transferability of results to other experiments. 3DoF captures users' rotational movement, while 6DoF additionally captures translational movement, allowing us to assess the practicality of the audio presentation conditions (Section~\ref{sec:audio_conditions}).

\subsection{Implementation}

\label{sec:audio_conditions}
All three rendering conditions shared a common simulation layer in Unity, which maintained the source position and listener pose in world space and applied the same logarithmic distance attenuation with a minimum distance of $1.5$~m. Within this radius, the source played at full gain; beyond it, gain rolled off logarithmically with distance, producing motion parallax cues in the 6DoF task as translation toward or away from the source modulated loudness. In the 3DoF task, distance attenuation and translational tracking were both disabled, restricting the participant to rotational motion. A constant reverberation profile was applied across all conditions to control for room-acoustic cues. The two HRTF-based conditions were implemented through the Steam Audio plugin for Unity, which supports import of generic HRTF measurements in the SOFA file format.

\emph{Condition 1 --- \condStereo}:
The 2-Channel condition bypassed HRTF rendering entirely, using Unity's built-in stereo panner (\texttt{spatialBlend = 0}) with an equal-power panning law to map source azimuth to left--right level differences. Source azimuth was computed each frame relative to the listener in the horizontal plane. Because this condition applied no direction-dependent filtering, it functioned as the least spatially rich presentation in the study and conveyed only coarse horizontal direction through stereo level differences.

\emph{Condition 2 --- \condAzim}:
This condition used a modified version of the MIT KEMAR dataset~\cite{gardner1994kemar} in which all non-zero elevation measurements were programmatically removed prior to importing into Unity via the Steam Audio Plugin. The modified HRTF model leaves only the horizontal-plane measurements at $0^\circ$ elevation. This preserved full $360^\circ$ azimuth coverage along with the interaural time and level cues that support horizontal localization, but excluded the elevation-dependent spectral variation associated with vertical localization and some front--back disambiguation~\cite{wightman1989headphone,wightman1989validation}. The condition was designed as a contracted binaural rendering: full horizontal directional structure with reduced vertical spectral detail.

\emph{Condition 3 --- \condFull}:
This condition used the unmodified MIT KEMAR dataset~\cite{gardner1994kemar}, preserving the complete set of measured binaural impulse responses across both azimuth and elevation. This retained interaural time differences, interaural level differences, and the elevation-dependent spectral cues that contribute to vertical and front--back localization in headphone-based virtual audio~\cite{wightman1989headphone,wightman1989validation}, making \condFull the richest spatial rendering in the study. Because the KEMAR measurements were not individualized, this condition remained subject to the well-established limitations of non-individualized HRTFs, including reduced elevation accuracy and increased front--back ambiguity for some listeners~\cite{wenzel1993localization}.

\subsubsection{SoundFleXR Target Placement Regions}
\label{subsubsec:angular-regions}
To stress-test front--back confusion, we selected mirrored azimuth pairs based on the well-established cone-of-confusion phenomenon in spatial hearing, where front and rear directions can produce similar binaural cues, especially when elevation-related spectral cues are weak or non-individualized~\cite{wenzel1993localization,romigh2014isolating,mcanally2014headmovement}. We therefore included mirrored front--back azimuth pairs together with direct lateral control positions, and evaluated each at two elevation levels. We define the azimuth and elevation sets:
\[
\Theta =
\{\pm15^\circ,\pm30^\circ,\pm90^\circ,\pm150^\circ,\pm165^\circ\},
\]
\[
\Phi =
\{0^\circ, +30^\circ\}
\]
so that each azimuth is evaluated both in the horizontal plane and at a moderate positive elevation.

The full stimulus set contained 20 unique directions (10 azimuths $\times$ 2 elevations). To ensure uniform distribution across directions, we generated a random permutation of the Cartesian product $\Theta \times \Phi$ and presented each direction exactly once per block, yielding 20 trials with equal representation of all azimuths and both elevation levels.

\subsubsection{Voice Stimuli and Spectral Control}

We used two speech recordings, one from a male speaker and one from a female speaker, both resampled to 44.1~kHz and loudness-normalized to $-20$~dBFS RMS, with a true-peak ceiling of $-1$~dBFS. On each trial, one of the two recordings was randomly selected as the source signal, providing within-participant counterbalancing of voice across trials. Male and female voices differ systematically in long-term spectral shape and fundamental frequency, with male speech typically carrying relatively more low-frequency energy and female speech typically carrying relatively more high-frequency energy~\cite{byrne1994ltass}.

Speech spectra can influence spatial details perceived, resulting in a potential confounding variable. Since the source spectrum can interact with the spectral cues listeners use for localization, including elevation-dependent pinna cues, we included both voices so that the task would not be implicitly tuned to only one portion of the speech spectrum~\cite{hebrank1974spectral,hofman2003binaural,alvespinto2014notches}. This was especially important given that some of the cues most relevant for vertical and front--back disambiguation are associated with high-frequency spectral structure, often in roughly the 5--10~kHz region~\cite{hebrank1974spectral,zonooz2019spectral, byrne1994ltass,kolarik2016distance}.

To reduce the differences between the two recordings, both signals were normalized to a target of $-20$~dBFS RMS with a true peak ceiling of $-1$~dBFS, bringing the inter-speaker level difference to below 1~dB. We also applied the same equalization chain to both signals before normalization, including an 80~Hz high-pass filter, a broadband presence boost from 2--5~kHz, and a high-frequency shelf above 8~kHz, so that the recordings were more perceptually consistent while still preserving the frequency range most relevant for spectral localization cues~\cite{hebrank1974spectral,alvespinto2014notches}. Spectral verification confirmed that the 4--6~kHz band remained intact after processing.

\begin{figure}[!htbp]
\centering
\includegraphics[width=0.75\linewidth]{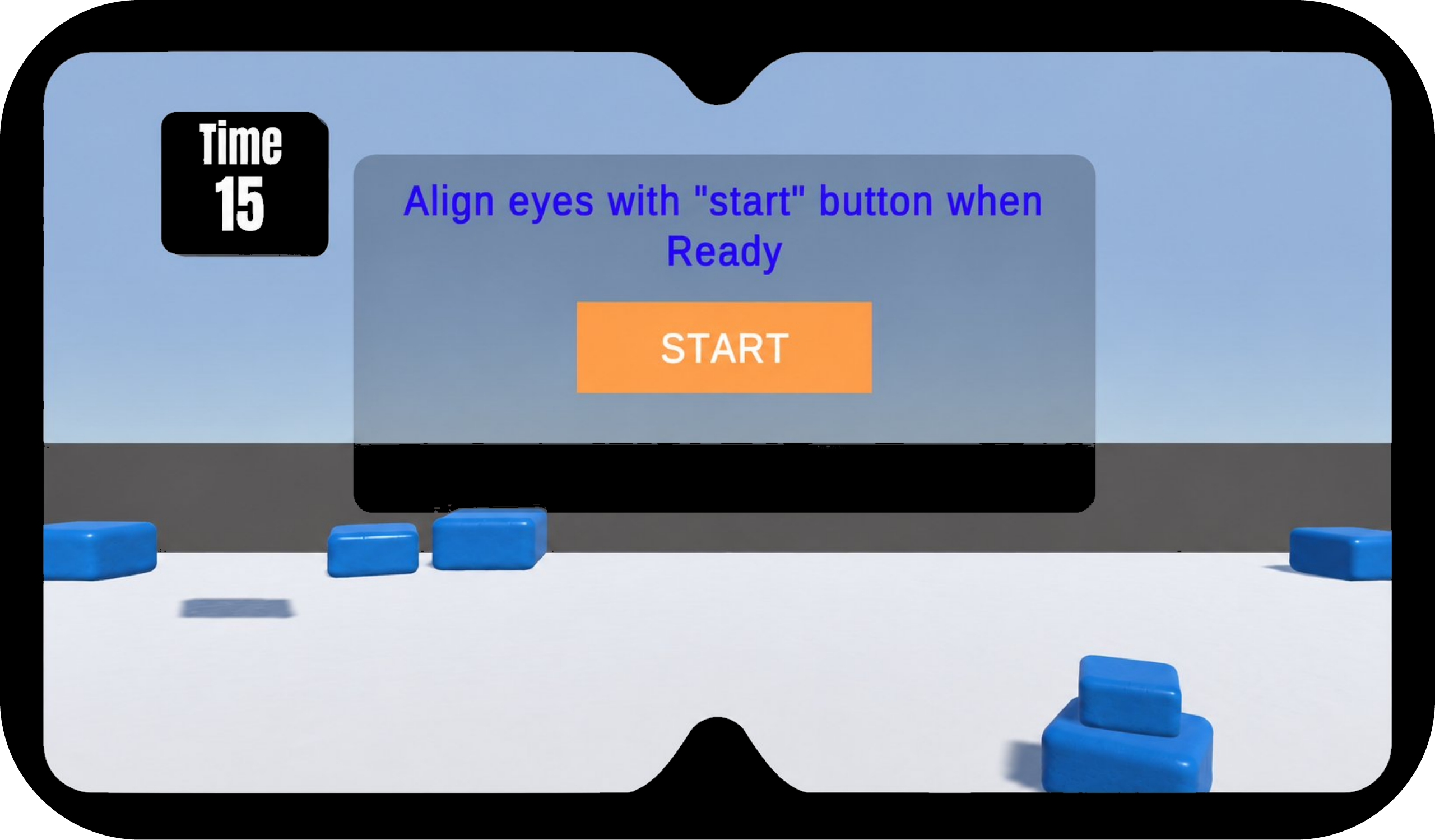}
\caption[SoundFleXR start menu]{Start menu shown in the AR passthrough environment at the beginning of each trial. Participants aligned their gaze with the START button to calibrate their head pose with the source spawn coordinate system, then pressed a button on the PlayStation~VR2 controller to begin the 15-second guess window. The trial timer is visible in the upper-left corner.}
\label{fig:flexr-ui}
\end{figure}

\subsubsection{Conforming Perceived Locations} To select a perceived audio location, participants use a visible \emph{Spatial Cursor} at the endpoint of a ray projected from their head pose. We use head-directed selection because controller-based pointing requires participants to first orient toward the perceived sound and then separately align the controller ray with that location. This additional movement can introduce pointing errors unrelated to auditory localization and to the general human errors observed in XR research \cite{valzolgher2020impact,cho2024auptimize,hayatpur2019planeraypoint,shan2024work}. The \emph{Spatial Cursor} followed HMD rotation and remained constrained to the boundary of the upper hemispherical collider delimiting the workspace. When they were satisfied with their aim, participants confirmed the guess with a button press on the PSVR2 controllers; on confirmation, the \emph{Spatial Cursor} ray endpoint was written to a CSV file saved in a container on the Apple Vision Pro. This head-aimed, button-committed workflow matches the spirit of head-based pointing responses common in spatial audio localization work~\cite{cho2024auptimize,berger2018generic,valzolgher2020impact}.

\section{Evaluation}

To answer RQ1 and RQ2 (Section~\ref{sec:intro}), we conducted a user study to assess whether audio presentations of varying dimensionality were non-inferior to one another in angular localization performance and user experience.

\subsection{Experimental Design}

The within-subjects experiment followed a $2 \times 3$ factorial design, with task DoF (3DoF and 6DoF) and audio presentation (\condStereo, \condAzim, and \condFull) as within-participant factors. To answer RQ1, we complemented the $2 \times 3$ factorial design with planned task-specific $1 \times 3$ analyses that assessed both directional non-inferiority and bidirectional equivalence within each task DoF (i.e., pairwise audio-presentation contrasts were also investigated separately for the 3DoF and 6DoF tasks).

\subsubsection{Tasks}
In the 3DoF task, participants remained seated in a swivel chair at the center of a physical space defined by a circle with a 3 m diameter. This constrained translational movement, so participants primarily used head and body rotation to localize the virtual sound source relative to themselves. In the 6DoF task, participants were free to move within the same physically defined workspace, allowing them to use both rotational and translational movement. This allowed us to compare localization performance when the task was constrained to orientation versus when participants could also use embodied movement strategies.

To support the interpretability of both tasks, user movement and source placement were restricted to the same 3 m circular workspace in both the physical room and the Unity scene. In the physical lab, floor markings defined the circle participants could not pass. In Unity, this same boundary was mirrored using a hemispherical collider centered on the participant's calibrated starting position.\footnote{The hemispherical collider in Unity is measured in \emph{game units}, which are translated to meters once ported on an HMD.} Target audio sources were always placed on the edges of the hemisphere at a $1.5$ m Euclidean distance from the HMD position. Audio sources spawned at the target placement regions detailed in Section~\ref{subsubsec:angular-regions}, so each block contained one randomized pass through all 20 azimuth-elevation combinations.

\subsection{Procedure}
Before beginning the formal trials, participants completed a training period using a randomized task DoF and audio presentation condition. During training, participants made 10 practice guesses, which pilot testing indicated were sufficient to understand the localization task, the START menu, and the selection interface. The experimenter instructed participants to listen for directional cues, aim with the \emph{Spatial Cursor}, then confirm with the PSVR2 controller button when they believed they faced the sound.

During both training and formal trials, participants were not shown the virtual sound source while making their estimate. Each trial began with the participant aligning their gaze with the START button and pressing a PSVR2 controller button. This alignment step registered the participant's head pose with the source spawn coordinate system, so the azimuth and elevation of each target direction were defined relative to the participant's calibrated starting pose. After the trial began, the system rendered a looping voice recording saying \enquote{hey look over here} from one of the target directions. The voice was randomly selected from recordings of one male and one female on each trial, which helped ensure that the findings were not tied to the spectral characteristics of a single voice.

The experiment included two localization tasks (3DoF and 6DoF) and three audio presentation conditions (\condStereo, \condAzim, and \condFull). Each participant completed one block for each task-by-audio condition combination, with 20 localization trials per block, for a total of 120 trials. The active localization portion took up to approximately 40 minutes, excluding questionnaires and breaks.

Participants had 15 seconds to make each localization guess. We used this time window to allow participants enough time to explore head movement strategies without permitting unbounded exploration that could reduce rendering-driven differences across tasks. While hearing the voice, they repeated the same aim-then-confirm procedure. After they confirmed their estimate, the true source location became visible as feedback before the next trial. The feedback appeared as a color-coded virtual head positioned at the true source location, with color indicating the audio presentation condition: \condStereo (orange), \condAzim (yellow), and \condFull (blue).

All participants completed both task conditions and all three audio presentation conditions. The order of task conditions and audio presentation blocks was counterbalanced across participants to reduce order effects. Target directions were randomly ordered within each block, so each target direction was represented equally across the azimuth and elevation combinations.

After completing task 1 within each posture condition (sitting or standing), participants answered the six NASA-TLX workload questions on 1--7 scales. We administered NASA-TLX once per posture condition rather than after every trial to avoid interrupting the series of localization tasks. These responses provided contextual workload measures at the audio-condition level (i.e., \condFull, \condAzim, and \condStereo), while the primary analyses reported below focus on localization error and qualitative feedback.

Upon completion of both tasks, participants completed a qualitative questionnaire informed by Endsley-style AR design heuristics~\cite{endsley2017augmented}. The questionnaire aimed to identify the most salient properties across audio presentations, task, localization, and user strategies---that is, the features prominent enough for participants to surface unprompted. That emphasis complements the quantitative data by capturing what participants attended to and made sense of during the task. Using participant reflections as qualitative signals for the most prominent experimental properties is consistent with procedures that assess user experience with spatial sound~\cite{poeschl2013integration,di2025audiovisual,woodard2025audiomixr}.

\subsection{Participants}

We recruited 18 participants (11 male, 7 female; mean age: 25.3 years) from a local university for the study. The sample size was informed by comparable spatial audio localization studies, which recruited 7--18 participants per condition across between- and within-subject designs~\cite{cho2024auptimize,ogawa2024exploring,aggiusvella2020comparison,gilberto2025virtual,cagiltay2026audio}. Participants reported a mean prior XR experience rating of 2.89 on a 1--7 scale, where 1 indicated ``\emph{no experience}'' and 7 indicated ``\emph{expert}.'' None reported any known hearing impairments in the preliminary questionnaire. All participants provided written consent and received a \$30 gift card as compensation.

\subsection{Apparatus}
\label{sec:soundflexr-system}
 \emph{SoundFleXR} is implemented as an augmented reality application so that participants could see the surrounding physical environment alongside the virtual UI elements (Figure~\ref{fig:flexr-ui}). We chose AR passthrough rather than a fully virtual environment to reduce the risk of disorientation and to keep the task perceptually grounded in the room participants were physically in, which is consistent with prior work on XR audio design where digital content is bidirectionally coupled to the physical context~\cite{woodard2025audiomixr,woodard2025cam, yang2019audio,gupta2022augmented}.

Participants wore an \textbf{Apple Vision Pro} as the HMD, which delivered that passthrough view and supplied head pose for aligning virtual content with the room---including the on-floor circle with a 3~m diameter used in both tasks---for rendering the aiming cue described below, and for logging each confirmed response direction. We presented audio over \textbf{Bose QuietComfort} closed-back headphones driven by our renderer. Consumer gaming headsets often ship with proprietary surround or spatialization that would sit outside our software stack, whereas this arrangement kept the listening path centered on our two-channel binaural output; \textbf{headphone spatial-audio or immersive listening modes were disabled when available} so that localization reflected our experimental presentations rather than additional manufacturer-side spatialization. Participants held \textbf{PlayStation~VR2 (PSVR2)} controllers and confirmed perceived location with a controller button. We preferred physical buttons to Vision~Pro hand pinch as the commit gesture because a discrete press is easy to standardize across hand sizes and dexterity and because it separates localization judgments from pinch-detection variability, which keeps trial-to-trial input more comparable across the study sample.

\begin{comment}
    \section{User Evaluation}
The experiment aims to characterize the varying configurations of \emph{two levels of interaction DoF} and \emph{rendering fidelity/three audio presentations} and to provide design lessons centered on task objectives.
\end{comment}
\subsection{Angular Margins}

For this study, the non-inferiority margins of interest were
\(5^\circ\), \(10^\circ\), \(15^\circ\), \(20^\circ\), and
\(30^\circ\). Prior work demonstrates that \(360^\circ\) angular
localization varies across spatial regions, with reported thresholds
and errors of approximately \(4^\circ\)--\(6^\circ\) for frontal and
rear sounds and \(20^\circ\)--\(22^\circ\) for sounds presented near
the listener's sides
\cite{aggiusvella2020comparison,valzolgher2020impact}.
Research specific to XR audio localization has also reported
localization errors and used angular segments around
\(10^\circ\)--\(15^\circ\)
\cite{cho2024auptimize,ogawa2024exploring}. The \(15^\circ\) and
\(30^\circ\) margins additionally outline the horizontal and vertical
angular differences included in our target placements (Section \ref{subsubsec:angular-regions}). We use these
margins to evaluate a range of localization precision rather than
treating any value as a universal perceptual limit.

\subsection{Analysis}

The goal of evaluating audio presentations is to determine when a simplified representation can function similarly to an HRTF-based spatial representation. We do this by performing non-inferiority (NI) tests across combinations of \condStereo, \condAzim, and \condFull in both 3DoF and 6DoF localization tasks.

Rather than treating \condFull as the baseline \enquote{best} rendering, we measure whether lower-dimensional renderings preserve localization accuracy relative to higher-dimensional renderings using mean-error non-inferiority tests and, separately, trial success rates computed under several angular tolerances (Appendix \ref{tab:ni-success-summary}). Together, these analyses specify when one representation can substitute for another in average error and in how often errors fall below an angular error threshold. When a lower-dimensional rendering passes the relevant criterion, it suggests that the simpler representation may be used in applications with similar spatial demands.

 In addition to comparing overall performance, we examine which spatial aspects are preserved or lost as audio dimensions are reduced. We therefore interpret the results not simply as better or worse, but in terms of how well each representation matches the task requirements.
For example, if \condAzim is non-inferior to \condFull for azimuth error but not elevation error, then \condAzim may be suitable for tasks requiring users to distinguish horizontal source locations, but less suitable for tasks requiring vertical discrimination. In this case, the reduced representation preserves the task-relevant azimuth dimension while omitting the auditory dimension, which may only matter in other task configurations.

The same logic applies when comparing \condStereo with the other representations. If \condStereo is only supported as a substitute under looser angular tolerances than \condAzim or \condFull, then it may provide good-enough directionality, but may be less appropriate for discriminating between closely spaced spatial locations.
Thus, the main focus of these analyses is to identify when reducing the number of audio dimensions does not produce a meaningful loss of task-relevant localization information.

\section{Results}
\label{sec:soundflexr-results}

We organize the results in two parts. First, we test within-task non-inferiority across five angular tolerances used for success-rate summaries.
($5^\circ$, $10^\circ$, $15^\circ$, $20^\circ$, $30^\circ$) and three error metrics
(azimuth, elevation, total angular). Azimuth error measures horizontal deviation, elevation error measures vertical deviation, and total angular error measures the combined angular separation between the true source direction and the participant's guess. Then, we conclude the results with a categorization of participant feedback to complement the quantitative results.

Table~\ref{tab:raw-angular-error-summary} reports the raw mean angular errors for each task and rendering condition. \condAzim and \condFull are closely matched for total angular error in both 3DoF and 6DoF, while \condStereo has larger total and azimuth errors. Elevation error remains similar across conditions, reinforcing that the main separation between renderings is driven more by horizontal localization than by vertical localization.

NASA-TLX results demonstrated that all conditions were non-inferior to each other across all NASA-TLX metrics — mental, physical, and temporal demands, as well as effort and frustration. We report these results in Appendix \ref{app:rawtlx}.

\begin{table*}[t]
\centering
\small
\caption{Raw mean angular error by task and audio presentation. Values are in degrees.}
\label{tab:raw-angular-error-summary}
\begin{tabular}{llrrr}
\toprule
Metric & Task & \condStereo & \condAzim & \condFull \\
\midrule
Total angular error & 3DoF & 33.5 & 28.3 & 28.2 \\
Total angular error & 6DoF & 41.9 & 36.6 & 36.4 \\
Azimuth error & 3DoF & 30.3 & 21.7 & 22.6 \\
Azimuth error & 6DoF & 32.8 & 26.6 & 28.7 \\
Elevation error & 3DoF & 15.6 & 16.1 & 16.2 \\
Elevation error & 6DoF & 16.6 & 17.3 & 16.0 \\
\bottomrule
\end{tabular}
\end{table*}

\subsection{Within-Task Non-Inferiority}

We tested whether the simplified presentations (\condStereo and \condAzim) were non-inferior to \condFull with respect to azimuth, elevation, and total angular error. For each NI margin ($\Delta \in {5^\circ,10^\circ,15^\circ,20^\circ,30^\circ}$), we conducted paired one-sided (t)-tests on participant mean error. These tests ask whether the \condStereo and \condAzim conditions produce less than ($\Delta$) degrees of angular mean error compared to \condFull. Since the NI margin changes, the resulting (p)-values may also change across different angular thresholds. 

We separately calculated success rates at each angular threshold ($\Delta$) to assess the proportion of target guesses within these margins with respect to the ground-truth target location, thereby revealing underlying consistency patterns that may not be visible in mean-error patterns alone.

Figures~\ref{fig:ni-success-azimuth-3dof}--\ref{fig:ni-success-total-6dof} report the mean-error NI (p)-values on the left and the paired success-rate differences relative to \condFull, with forest plots detailing the success rates/proportions. The full report, including all success-rate comparisons we study and their CIs, can be found in Appendix~\ref{app:ni-results} and Table~\ref{tab:ni-success-summary}. Additionally, we report the raw angular errors in Table~\ref{tab:raw-angular-error-summary} to guide interpretation of all aforementioned results.

\subsubsection{Elevation Error}

Every pairwise comparison met the mean-error non-inferiority criterion for elevation in both the 3DoF and 6DoF tasks. Paired mean-error NI tests indicated mutually non-inferiority at every angular tolerance tested, including the strictest \(5^\circ\) cut. The mean absolute differences between presentations were on the order of $0$--$1^\circ$, well inside typical tolerances, indicating that elevation localization accuracy was statistically equivalent across rendering conditions on both average error and success rates. The spectral cues afforded by the \condAzim and \condFull presentations did not translate into a measurable elevation advantage over \condStereo rendering in this task. The elevation error results are detailed in Appendix~\ref{app:ni-results}, Table~\ref{tab:ni-success-summary}.

\subsubsection{Azimuth Error}

\begin{figure*}[t]
    \centering
    \includegraphics[width=\textwidth]{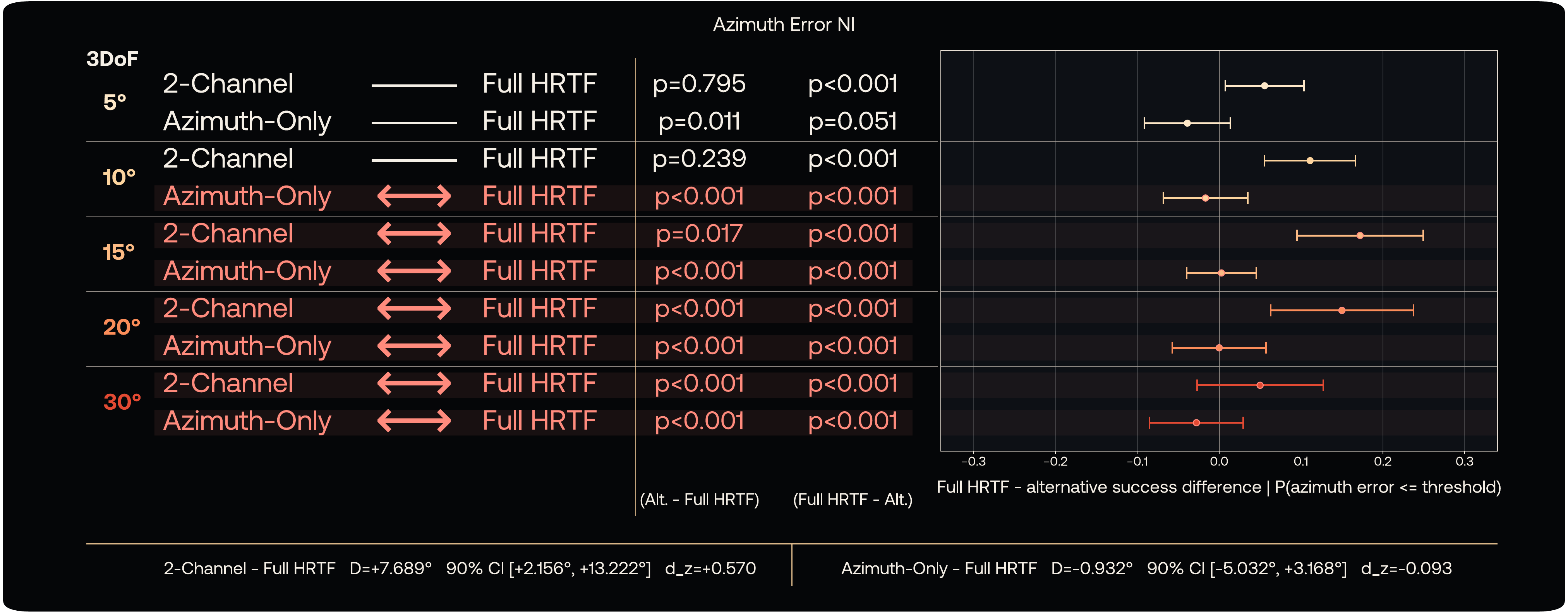}
    \caption[Azimuth non-inferiority success in 3DoF]{Within-task non-inferiority results for azimuth error in the 3DoF task across five angular thresholds ($5^\circ$, $10^\circ$, $15^\circ$, $20^\circ$, $30^\circ$). \emph{Left:} mean-error NI annotations (\(p\)-values). \emph{Right:} paired differences in success rate between \condFull and each alternative (points with intervals). \\

    }
    \label{fig:ni-success-azimuth-3dof}
\end{figure*}

The azimuth results show a stronger dependence on both angular tolerance and task condition. In the 3DoF task, mean-error NI tests also display mutual non-inferiority of \condAzim and \condFull at \(10^\circ\) and larger tolerances. In contrast, only the \condAzim $\rightarrow$ \condFull direction was supported at the strictest \(5^\circ\) tolerance (Figure~\ref{fig:ni-success-azimuth-3dof}). The mean-error NI tests also indicated mutual non-inferiority of \condStereo and \condFull beginning at the \(15^\circ\) tolerance, but not at \(5^\circ\) or \(10^\circ\). Thus, in the 3DoF task, all presentations converged at coarser azimuth tolerances, while the strictest tolerances separated \condStereo from the HRTF-based presentations.

\begin{figure*}[t]
    \centering
    \includegraphics[width=\textwidth]{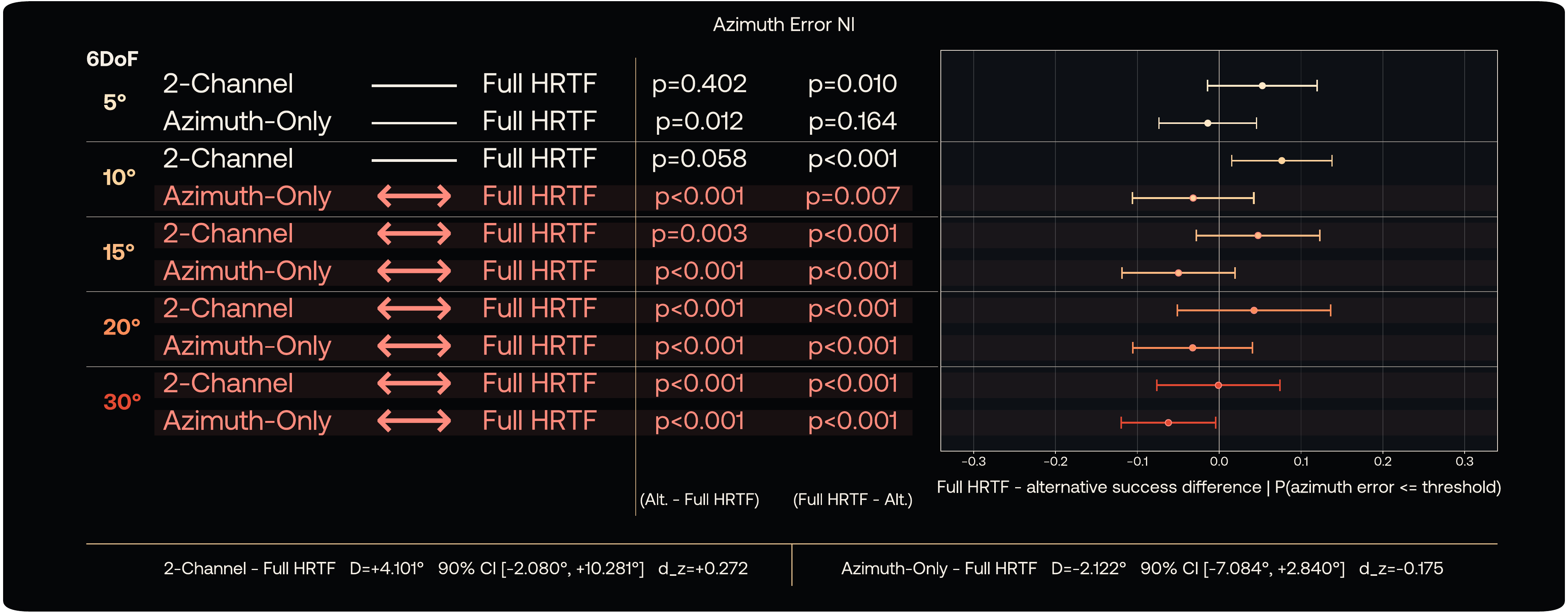}
    \caption[Azimuth non-inferiority success in 6DoF]{Within-task non-inferiority results for azimuth error in the 6DoF task across five angular thresholds ($5^\circ$, $10^\circ$, $15^\circ$, $20^\circ$, $30^\circ$). \emph{Left:} mean-error NI annotations (\(p\)-values). \emph{Right:} paired differences in success rate between \condFull and each alternative (points with intervals).}
    \label{fig:ni-success-azimuth-6dof}
\end{figure*}

The 6DoF azimuth results followed a similar pattern, but the separation between \condStereo and the HRTF-based presentations persisted slightly longer (Figure~\ref{fig:ni-success-azimuth-6dof}). The mean-error NI tests when comparing \condAzim and \condFull at \(10^\circ\) and larger tolerances displayed non-inferiority, with \condAzim $\rightarrow$ \condFull supported at \(5^\circ\). \condStereo and \condFull were found to be non-inferior to each other at \(15^\circ\) and larger tolerances, while the \(10^\circ\) comparison narrowly failed the success-rate non-inferiority criterion. This suggests that \condStereo can preserve broad azimuthal direction at coarser tolerances, but becomes less reliable when the task requires tighter azimuthal agreement with the HRTF-based reference.

\subsubsection{Total Angular Error}

\begin{figure*}[t]
    \centering
    \includegraphics[width=\textwidth]{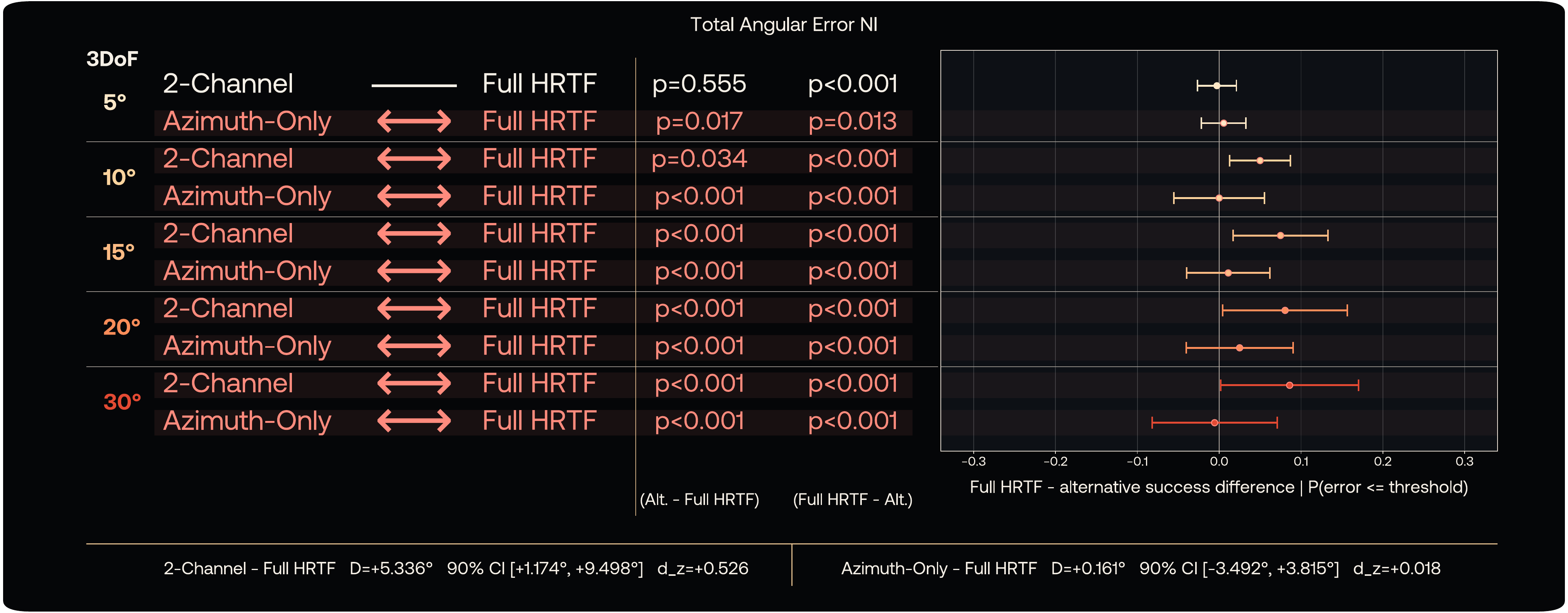}
    \caption[Total angular non-inferiority success in 3DoF]{Within-task non-inferiority results for total angular error in the 3DoF task across five angular thresholds ($5^\circ$, $10^\circ$, $15^\circ$, $20^\circ$, $30^\circ$). \emph{Left:} mean-error NI annotations (\(p\)-values). \emph{Right:} paired differences in success rate between \condFull and each alternative (points with intervals).}
    \label{fig:ni-success-total-3dof}
\end{figure*}

The total angular error results show that \condAzim and \condFull were highly comparable across both task conditions. In the 3DoF task,  \condAzim and \condFull were non-inferior to each other at every tolerance tested, including the strictest $5^\circ$ cut (Figure~\ref{fig:ni-success-total-3dof}). Our NI tests for the 3DoF task also found  \condStereo and \condFull to be non-inferior to each other at $10^\circ$ and larger tolerances, but not at $5^\circ$. This indicates that, for total angular error in the 3DoF task, \condStereo becomes comparable to \condFull once the acceptable angular tolerance is relaxed beyond the strictest cut.

\begin{figure*}[t]
    \centering
    \includegraphics[width=\textwidth]{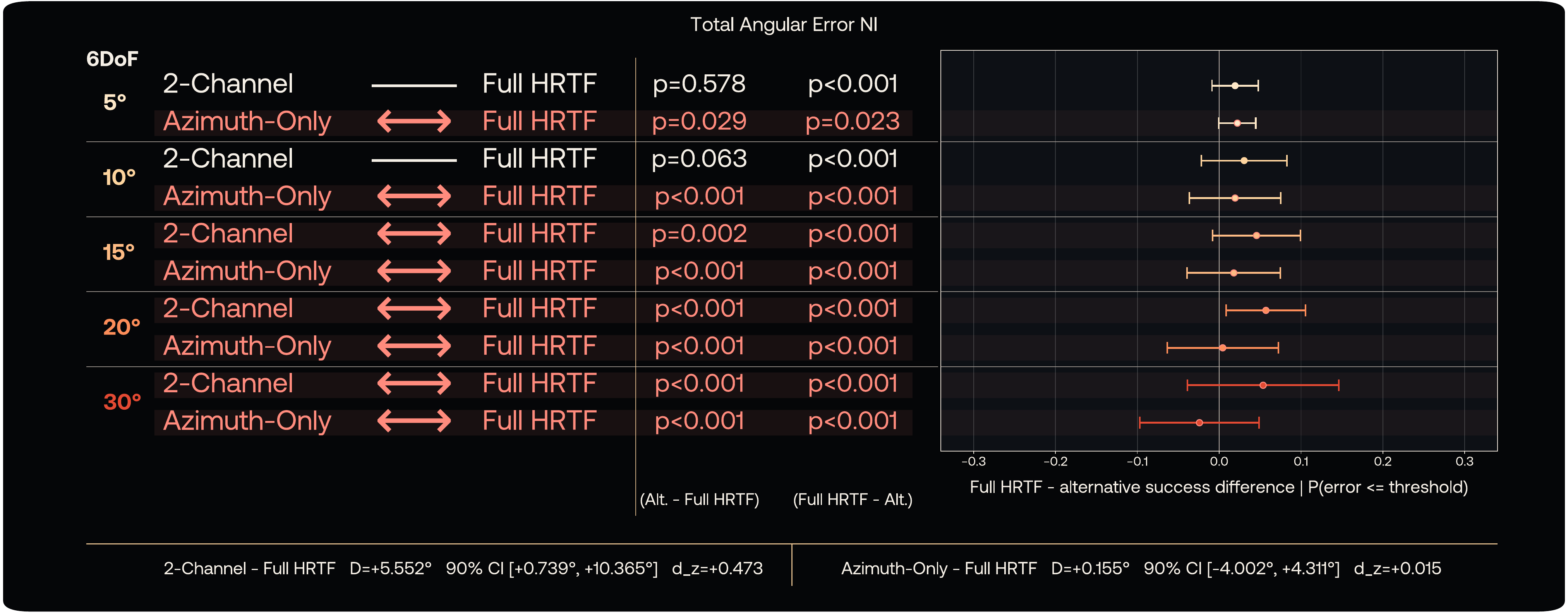}
    \caption[Total angular non-inferiority success in 6DoF]{Within-task non-inferiority results for total angular error in the 6DoF task across five angular thresholds ($5^\circ$, $10^\circ$, $15^\circ$, $20^\circ$, $30^\circ$). \emph{Left:} mean-error NI annotations (\(p\)-values). \emph{Right:} paired differences in success rate between \condFull and each alternative (points with intervals).}
    \label{fig:ni-success-total-6dof}
\end{figure*}

The 6DoF total angular error results were similar but slightly more conservative for \condStereo (Figure~\ref{fig:ni-success-total-6dof}). Again, our analysis revealed mutual non-inferiority of \condAzim and \condFull at every tolerance tested, including the strictest $5^\circ$ cut. Similar to the 6DoF azimuth analysis, \condStereo and \condFull were mutually non-inferior at $15^\circ$ and larger tolerances, but not at $5^\circ$ or $10^\circ$. Thus, \condStereo required a coarser tolerance to be considered comparable in the 6DoF task, while \condAzim remained closely aligned with \condFull across the full set of tolerances.

\begin{figure*}[t]
\centering
\includegraphics[width=0.8\textwidth]{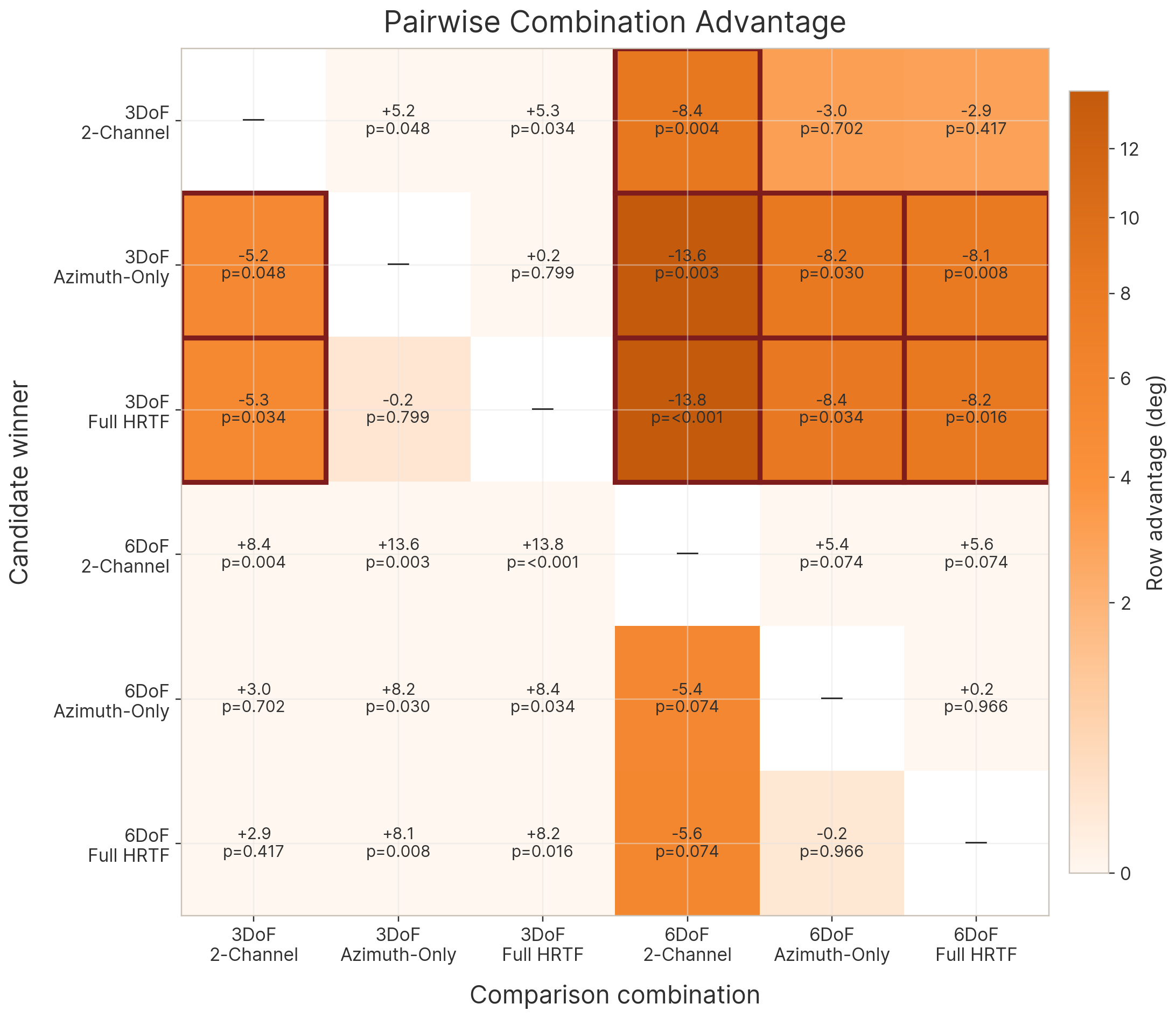}
\caption[Pairwise advantage combination heatmap]{Pairwise advantage heatmap over the six combinations of task DoF (3DoF vs.\ 6DoF) and audio presentation (\condStereo, \condAzim, \condFull). Each off-diagonal cell gives the signed mean advantage in degrees for the row combination relative to the column combination, with a $p$-value; shading emphasizes larger-magnitude advantages in this color mapping.}
\label{fig:pairwise-advantage-combination-heatmap}
\end{figure*}

\subsection{Pairwise Advantage Analysis}

We examined pairwise advantage among the six cells defined by task DoF (3DoF vs.\ 6DoF) and audio presentation (\condStereo, \condAzim, \condFull). Figure~\ref{fig:pairwise-advantage-combination-heatmap} reports signed mean advantages in degrees for each row cell relative to each column cell, with $p$-values and color emphasis. Non-inferiority analyses above test substitution against \condFull on mean error and success within the specified angular thresholds; the heatmap adds a compact view of mean-error \emph{pairwise} dominance across the full factorial grid.

\subsection{Qualitative Feedback}

Each participant provided open-ended written feedback after the
experimental tasks. Responses were organized into the following categories.

\subsubsection{Elevation Localization}

Eight of eighteen participants 
reported difficulty resolving the vertical position of sources.
Comments described the horizontal axis as comparatively easy while
up--down remained ambiguous. For example, P8 wrote that \enquote{horizontal
direction[s] were detectable, but vertical direction was hard,} while
P17 summarized the same pattern as \enquote{left/right was somewhat clear but
up/down not clear.} Several participants also reported defaulting to
the midpoint of the perceived elevation range when uncertain. The
difficulty was distributed across all three rendering conditions rather
than concentrated in any one. This pattern is consistent with the
within-task non-inferiority result on elevation error, in which all presentations were
statistically equivalent at every margin tested: the additional
spectral cues in \condAzim and \condFull did not produce a perceived
elevation advantage over \condStereo.

\subsubsection{Task Demands of 6DoF}

Five participants compared the seated 3DoF
and standing 6DoF tasks. Standing was more often described as natural
or immersive, while seated rotation was reported as more convenient
and dimensionally constrained. P11 described standing as \enquote{more
natural} and \enquote{more immersive,} while P16 noted that, when seated,
they \enquote{cannot move around to see if one direction is louder.} These
accounts align with the consistent $\sim$$8^\circ$ accuracy cost of
6DoF observed across all rendering conditions (Section~4.1), and
suggest that the translational DoF imposes a localization
cost that subjective ease does not predict.

\subsubsection{Rendering Comparisons}

Nine participants explicitly compared the
three presentations. \condAzim was most frequently described as easiest
to localize or most natural, \condFull received mixed reports of
clarity versus difficulty, and \condStereo drew the most ambiguous
responses. P1 wrote that \enquote{Yellow (\condAzim) felt the most natural} and was
\enquote{the easiest} direction to detect. In contrast, P11 described
\condStereo as providing only \enquote{a vague sense} of direction but not
enough to \enquote{pinpoint the precise location}.This fragmentary,
non-systematic preference structure is consistent with the modest
separation in threshold-based success rates and with the within-task
non-inferiority results indicating that the presentations are statistically
equivalent at margins of $15^\circ$ and larger.
\subsubsection{Additional Observations}

The remaining themes were tangential to the quantitative hypotheses
and are summarized for completeness. Fifteen participants
described localization strategies, most commonly rotating the head
until both ears matched in volume and walking toward the perceived
source. Five participants reported front/back confusions, most often
for sources directly behind the listener. Eight commented on the
perceived stability or anchoring of sources, three on voice
characteristics and the use of ambient cues, four on suspected
technical artifacts such as audible panning, and two on order
effects as a potential source of bias in their condition preferences.

\section{Discussion}

We interpret our results and address our research questions by providing the following design lessons and application scenarios. Before turning to those lessons, we first introduce the framing through which we interpret them.

\subsection{Design Lesson \#1 --- Source Localization Requires Tight Angular Tolerances}

Under the strictest success-rate angular tolerances ($5^{\circ}$ and $10^{\circ}$), \condStereo often could not be supported as non-inferior to the \condFull reference condition, particularly when 6DoF translation was available (Figures~\ref{fig:ni-success-total-6dof} and~\ref{fig:ni-success-azimuth-6dof}). The design question is not only whether a reduced presentation lowers mean error relative to \condFull, but whether it remains acceptable at the angular tolerance the task requires. The remainder of this subsection outlines the data supporting this finding. It motivates why tight angular tolerances matter when XR tasks require a substitute audio presentation to preserve angular localization performance relative to an HRTF-based reference.

\subsubsection{Significant Findings}

Across the success-rate non-inferiority plots, \condAzim was the substitute audio presentation most consistently supported as non-inferior to \condFull under tighter angular tolerances. In the 3DoF task, this pattern appears for both total angular error and azimuth error (Figures~\ref{fig:ni-success-total-3dof} and~\ref{fig:ni-success-azimuth-3dof}). In the 6DoF task, \condAzim likewise remains closer to the \condFull reference than \condStereo across the corresponding total-angular and azimuth-error outcomes (Figures~\ref{fig:ni-success-total-6dof} and~\ref{fig:ni-success-azimuth-6dof}). This does not imply that either condition achieved $5^{\circ}$ average localization accuracy. Instead, the tested angular tolerances indicate where a designer could reduce the dimensions of audio feedback without meaningfully degrading success relative to \condFull.

The raw mean errors in Table~\ref{tab:raw-angular-error-summary} provide the descriptive version of the same pattern: \condAzim and \condFull remain close for total angular error, while \condStereo is higher for total and azimuth error. The pairwise advantage results (Figure~\ref{fig:pairwise-advantage-combination-heatmap}) indicate that these differences are task- and presentation-dependent rather than universally favoring one rendering.

\condStereo, by contrast, was not consistently supported as non-inferior to \condFull for either total angular error or azimuth error in the 3DoF and 6DoF tasks (Figures~\ref{fig:ni-success-total-3dof},~\ref{fig:ni-success-total-6dof},~\ref{fig:ni-success-azimuth-3dof}, and~\ref{fig:ni-success-azimuth-6dof}). In other words, paired success rates under strict angular tolerances did not establish that \condStereo matched \condFull often enough relative to the Full HRTF baseline. This outcome suggests that audio with spatial properties comparable to \condStereo is less suitable when the task requires an audio presentation to preserve HRTF-like localization performance under a tight angular tolerance. This interpretation is consistent with participant feedback, in which \condStereo was sometimes described as providing only \enquote{a vague sense} of direction, but not enough to \enquote{pinpoint the precise location}.

The pattern observed at the $5^{\circ}$ tolerance largely extends to the $10^{\circ}$ tolerance. \condStereo appears closer to the success-rate non-inferiority criterion for some total-angular comparisons. Still, it remains less supported for azimuth error, with this gap becoming more pronounced once 6DoF translation is included (Figures~\ref{fig:ni-success-azimuth-3dof} and~\ref{fig:ni-success-azimuth-6dof}). This disparity likely stems from the additional direction-dependent filtering available in \condFull and \condAzim, which supports spatial updates that are not available in \condStereo. \condStereo shares the same distance attenuation and reverberation implementation as the other conditions, but the HRTF-based conditions provide direction-dependent spectral filtering and head-shadow cues that update as the listener moves relative to the source~\cite{kim2001parallax, brungart1999parallax,zotkin2004renderingLocalized,kapralos2008dimensionalityReducedHRTFs}. These spatial details are primarily accessed and amplified through 6DoF movement, suggesting that a rendering configuration with the dimensions of \condStereo could be a weaker substitute when the task allows translation and when success-rate non-inferiority to an HRTF-based reference must hold at $5^{\circ}$ or $10^{\circ}$ angular tolerances. Participant comments also suggest that movement was not uniformly beneficial, with one participant noting that when seated they \enquote{cannot move around to see if one direction is louder,} while other participants described standing as more natural or immersive.

\subsection{Design Lesson \#2 --- Simplifying Audio Feedback for Coarse Guidance}

A second pattern is that \condStereo appears to function more as a coarse directional baseline than as a full spatial localization interface. Participants could often orient to a general region with it, but it was less reliable when performance depended on resolving specific spatial dimensions within tighter angular margins. Across tested angular thresholds, \condStereo became more acceptable only as the tolerance became looser, suggesting that simpler renderings may be appropriate when the task does not require fine directional precision. For XR design, this suggests that \condStereo could serve as a simpler alternative to more complex rendering when the task only requires rough left-right awareness, ambient context, or coarse directional prompting. For example, a notification indicating that something happened to their left may not require elevation cues, individualized spectral filtering, or full 3D externalization.

Coarse guidance remains a meaningful XR design goal. Many audio cues do not need to specify an exact source location. They may instead direct attention, confirm that an action occurred, indicate that a state changed, or help users maintain awareness of events outside the field of view. Prior work has similarly shown that spatialized audio can support search, navigation, immersion, and awareness in XR, but that its value depends on the interaction context and the cues users need to act on~\cite{ruminski2015experimental,poeschl2013integration,di2025audiovisual,potter2022relativeImportance,kailas2021design}. In these cases, simpler rendering may be beneficial because it avoids adding spatial detail that the user does not need to complete the task.

Thus, the broader lesson is not that \condStereo is generally equivalent to HRTF-based rendering. Rather, \condStereo marks one end of the design space where audio provides lightweight directional awareness without fully supporting precise spatial disambiguation. \condAzim occupies a middle ground by preserving horizontal spatial structure while omitting elevation-specific cues, and \condFull may be justified when the task requires full 3D angular interpretation at higher elevation angles since a standard HRTF should support monospectral cues that vertical localization relies on \cite{wenzel1993localization,wightman1989headphone}. For coarse guidance tasks, designers could simplify audio feedback when the reduced rendering still preserves the dimensions the user needs to act.

\subsection{Toward Task-Flexible Audio Rendering}

The main implication across these two lessons is not simply that high-precision XR tasks require more detailed audio rendering. Rather, the results suggest that complex audio presentations could be substituted with simpler ones when the reduced presentation preserves the spatial dimensions needed by the task. A task primarily structured around horizontal orientation, such as turning toward an off-screen notification, identifying which doorway a sound came from, or locating a speaker around the user, may not require the full set of elevation-dependent spectral cues provided by \condFull. In these cases, \condAzim can preserve the task-relevant horizontal structure while omitting dimensions unnecessary for the interaction.

The success-rate non-inferiority results suggest that \condAzim can remain close to \condFull under tight angular margins, particularly when the relevant task structure is primarily horizontal (Figures~\ref{fig:ni-success-azimuth-3dof} and~\ref{fig:ni-success-azimuth-6dof}). This does not mean that either rendering produced $5^{\circ}$ absolute localization accuracy. Rather, it means that \condAzim repeatedly remained acceptable relative to \condFull across tested angular tolerances. For horizontally structured tasks, this suggests that generic-HRTF azimuthal rendering could substitute for full generic-HRTF rendering when the user's goal is to resolve direction in the azimuthal plane. Participants also frequently identified \condAzim as the rendering that was \enquote{most easily localizable}, \enquote{most natural} or \enquote{the easiest} direction to detect. This suggests that the added spectral detail afforded by a generic full HRTF did not necessarily translate into a clearer experiential advantage when elevation was not central to the task.

This difference may be attributed to the absence of elevation cues in \condAzim, which may increase the perceptual clarity of azimuthal cues when the task does not require users to resolve height. Non-individualized HRTFs can introduce directional error relative to personalized or individualized HRTFs~\cite{wenzel1993localization, romigh2014isolating}, and this mismatch can conflict with users' own pinna anatomy. Recent work on numerical HRTF calculation further shows that individualized rendering remains an active technical direction rather than a solved default assumption~\cite{brinkmann2023recent}. \condAzim, which avoids making strong elevation claims, may sidestep this spectral mismatch and sound less perceptually cluttered, especially when the task is primarily organized around horizontal localization. This interpretation is consistent with the azimuth-error non-inferiority results, where \condAzim remains difficult to displace under tighter angular margins (Figures~\ref{fig:ni-success-azimuth-3dof} and~\ref{fig:ni-success-azimuth-6dof}). It also aligns with participants' broader comments that horizontal direction was more detectable than vertical direction, with one participant writing that \enquote{horizontal direction[s] were detectable, but vertical direction was hard.}

However, this substitution depends on the task's required dimensions. In tasks such as multi-shelf warehouse picking, aerial threat warnings, locating a speaker on a balcony, or distinguishing whether an alert came from above or below, elevation becomes part of the task structure. In these cases, \condFull may be harder to substitute because it contributes elevation information that \condAzim's constrained elevation rendering cannot provide––-our results suggest that angular elevation between ($0^\circ$) and ($+30^\circ$) may be negligible across conditions, but a study with more variation could characterize this further (Elevation results can be found in Appendix \ref{app:ni-results}). 

This dimensional framing also applies to tasks that are precise but not fully three-dimensional. For example, modifying a detailed CAD model in XR while restricting interaction to a horizontal work plane may require precise control, but not necessarily elevation-rich audio. In that case, \condAzim could be used in place of \condFull because the task demands precision in the horizontal plane rather than full 3D localization. Conversely, a less precise task may still require \condFull if the critical distinction is vertical, such as identifying whether a warning came from the floor, head height, or above the user. Thus, the relevant design question is not only how precise the task is, but which spatial dimensions the task requires users to resolve.

In sum, simpler audio presentations can substitute for more complex ones when they preserve the dimensions the task requires at the tolerance the task permits. \condAzim is a strong candidate as a substitute when the task is horizontally structured; \condFull becomes harder to replace when the task requires elevation discrimination or full 3D angular structure; and \condStereo may be sufficient when rough directional guidance or left-right awareness is sufficient. \condStereo's weaker support under tighter angular margins indicates a limit to how much spatial rendering can be simplified before the reduction becomes measurable in localization performance. Still, the broader contribution is dimensional: simpler renderings can be appropriate when they preserve the cues the task actually needs.

The results also suggest that richer rendering becomes harder to substitute as interaction expands beyond simple horizontal orientation. \condFull did not yield a broad timing advantage, but it was often perceived as more natural and spatially grounded, and participants described its benefits for movement, posture, and externalization. This suggests that richer spatial rendering may be most valuable not merely because it is more realistic, but because it can better support embodied interaction when users can move, reorient, and interpret a scene through multiple spatial cues~\cite{begault2001direct,mcanally2014headmovement,gulli2022spontaneous}.

\section{Limitations and Future Work}

Design lessons from this study are best suited for XR tasks in which audio dimensions can be clearly identified, such as embodied spatial control and static source localization. This study is less conclusive for dynamic, dense, long-duration, or highly ecological XR environments (e.g., crowded in-game environments with multiple audio sources being rendered simultaneously). \emph{SoundFleXR}, for example, tested static invisible sound sources rather than moving sources. A moving sound introduces additional task demands, including trajectory tracking, velocity estimation, prediction, and continuous spatial updating. Therefore, the results do not prove that the same rendering choices would work equally well for moving hazards, animated agents, dynamic alerts, or sources that travel through a scene.

Additionally, because the HRTF-based renderings used in this study relied on generic rather than personalized HRTFs, the results should not be interpreted as evidence that full individualized HRTFs are unnecessary. Begault et al. compared head tracking, reverberation, and individualized HRTFs for virtual speech perception, demonstrating that spatial perception depends on multiple interacting factors in both rendering and listeners~\cite {begault2001direct}. Work on numerical HRTF calculation and real-time acoustic field modeling further suggests that future systems may need richer perceptual and technical models than those tested under the controlled conditions here~\cite{brinkmann2023recent,lindau2014saqi,tong2025multimodal}.

\section{Conclusion}
This research supports the view that XR audio feedback should not be evaluated only by how realistically it reproduces physical listening. Results across the 3DoF and 6DoF localization experiments show that audio presentations may be interchangeable when the reduced presentation preserves the dimensions users need to perceive, resolve, and act upon. The non-inferiority results provide direct support for this task-dependent view of audio feedback dimensions. \condAzim was often a viable substitute for \condFull when users localized objects within relatively tight angular margins and when the task's relevant structure was primarily horizontal. By contrast, \condStereo was not consistently supported as a substitute for \condFull when localization required finer angular resolution below $5^\circ$ for 3DoF movement or below $10^\circ$ for 6DoF movement due to additional spatial cues. These findings suggest that the dimensionality of an audio presentation can be reduced without necessarily diminishing its functional value, provided the omitted dimensions are not critical to the task. Constrained audio representations are not simply lower-fidelity versions of complex renderings; they can function as strategic design choices that give designers flexibility to simplify, constrain, or extend audio feedback according to the tolerance geometry and spatial demands of a particular task.

\section*{Acknowledgments}
We sincerely thank all participants for contributing their time and effort for this study.

\bibliographystyle{IEEEtran}
\bibliography{arxiv}
\onecolumn
\clearpage
\raggedbottom
\setlength{\intextsep}{6pt}

\appendices
\section{Detailed Raw-TLX Non-Inferiority Results}
\label{app:rawtlx}

\begin{table}[H]
\centering
\caption{Non-inferiority tests for aggregate Raw TLX workload at
$\Delta=0.50$. Differences are calculated as
$D=\text{alternative presentation}-\text{\condFull}$, with higher values
indicating greater workload. Arrows indicate the supported NI direction;
$A\leftrightarrow B$ denotes mutual non-inferiority. NI $p$-value pairs are
reported in the comparison's left-to-right/right-to-left order. Mutual
non-inferiority is supported when the 90\% CI for $D$ lies entirely within
$[-0.50,0.50]$. Raw TLX is the unweighted mean of the six NASA-TLX items,
with Performance reverse-coded. Tests are unadjusted.}
\label{tab:tlx-ni}

\scriptsize
\setlength{\tabcolsep}{4pt}
\begin{tabular}{lllrlr}
\toprule
Metric & Task & Comparison & NI $p$ & TLXDiff & 90\% CI \\
\midrule
Raw TLX & 3DoF & \condStereo{} $\leftrightarrow$ \condFull{}
& $<.001/<.001$ & 0.000 & $[-0.188,\ 0.188]$ \\
& 3DoF & \condAzim{} $\leftrightarrow$ \condFull{}
& $.002/<.001$ & 0.037 & $[-0.213,\ 0.287]$ \\
& 6DoF & \condStereo{} $\leftrightarrow$ \condFull{}
& $<.001/<.001$ & 0.037 & $[-0.180,\ 0.254]$ \\
& 6DoF & \condAzim{} $\leftrightarrow$ \condFull{}
& $<.001/<.001$ & 0.000 & $[-0.227,\ 0.227]$ \\
\bottomrule
\end{tabular}
\end{table}

\section{Detailed Non-Inferiority Results}
\label{app:ni-results}

\begin{table}[H]
\centering
\caption{Mean-error non-inferiority (NI) and success-rate differences
across angular margins \(t\) for azimuth and total angular error.
NI \(p\) reports paired one-sided \(t\)-tests on participant mean error
with margin \(t^{\circ}\). The highlighted \(A \leftrightarrow B\) represent mutual NI, whereas a single \(\rightarrow \) implies the initial alternative condition comparison with \condFull did not achieve non-inferiority.
A trial succeeded when its error relative to the target (ground truth) was less than ($\Delta$). SuccessDiff is the
paired success-rate difference (Full HRTF minus the alternative), with
90\% CIs. }
\label{tab:ni-success-summary}

\fontsize{6.2pt}{6.7pt}\selectfont
\setlength{\tabcolsep}{3pt}

\begin{tabular}{@{}p{0.10\textwidth}llp{0.28\textwidth}
p{0.15\textwidth}rp{0.15\textwidth}@{}}
\toprule
Metric & Task & \(t\) & Comparison & NI p & SuccessDiff & 90\% CI \\
\midrule
Total angular & 3DoF & 5 & 2-Channel \(\rightarrow\) Full HRTF & \(p=0.555\) & \(-0.003\) & \([-0.027, 0.021]\) \\
\rowcolor{contribpurple}
Total angular & 3DoF & 5 & Azimuth-Only \(\leftrightarrow\) Full HRTF & \(p=0.017 / 0.013\) & \(0.006\) & \([-0.022, 0.033]\) \\
\rowcolor{contribpurple}
Total angular & 3DoF & 10 & 2-Channel \(\leftrightarrow\) Full HRTF & \(p=0.034 / <0.001\) & \(0.050\) & \([0.013, 0.087]\) \\
\rowcolor{contribpurple}
Total angular & 3DoF & 10 & Azimuth-Only \(\leftrightarrow\) Full HRTF & \(p<0.001 / <0.001\) & \(0.000\) & \([-0.055, 0.055]\) \\
\rowcolor{contribpurple}
Total angular & 3DoF & 15 & 2-Channel \(\leftrightarrow\) Full HRTF & \(p<0.001 / <0.001\) & \(0.075\) & \([0.017, 0.133]\) \\
\rowcolor{contribpurple}
Total angular & 3DoF & 15 & Azimuth-Only \(\leftrightarrow\) Full HRTF & \(p<0.001 / <0.001\) & \(0.011\) & \([-0.040, 0.062]\) \\
\rowcolor{contribpurple}
Total angular & 3DoF & 20 & 2-Channel \(\leftrightarrow\) Full HRTF & \(p<0.001 / <0.001\) & \(0.081\) & \([0.004, 0.157]\) \\
\rowcolor{contribpurple}
Total angular & 3DoF & 20 & Azimuth-Only \(\leftrightarrow\) Full HRTF & \(p<0.001 / <0.001\) & \(0.025\) & \([-0.040, 0.090]\) \\
\rowcolor{contribpurple}
Total angular & 3DoF & 30 & 2-Channel \(\leftrightarrow\) Full HRTF & \(p<0.001 / <0.001\) & \(0.086\) & \([0.002, 0.170]\) \\
\rowcolor{contribpurple}
Total angular & 3DoF & 30 & Azimuth-Only \(\leftrightarrow\) Full HRTF & \(p<0.001 / <0.001\) & \(-0.006\) & \([-0.082, 0.071]\) \\

Total angular & 6DoF & 5 & 2-Channel \(\rightarrow\) Full HRTF & \(p=0.578\) & \(0.019\) & \([-0.009, 0.048]\) \\
\rowcolor{contribpurple}
Total angular & 6DoF & 5 & Azimuth-Only \(\leftrightarrow\) Full HRTF & \(p=0.029 / 0.023\) & \(0.022\) & \([-0.000, 0.045]\) \\
Total angular & 6DoF & 10 & 2-Channel \(\rightarrow\) Full HRTF & \(p=0.063\) & \(0.031\) & \([-0.022, 0.083]\) \\
\rowcolor{contribpurple}
Total angular & 6DoF & 10 & Azimuth-Only \(\leftrightarrow\) Full HRTF & \(p<0.001 / <0.001\) & \(0.019\) & \([-0.036, 0.075]\) \\
\rowcolor{contribpurple}
Total angular & 6DoF & 15 & 2-Channel \(\leftrightarrow\) Full HRTF & \(p=0.002 / <0.001\) & \(0.046\) & \([-0.008, 0.100]\) \\
\rowcolor{contribpurple}
Total angular & 6DoF & 15 & Azimuth-Only \(\leftrightarrow\) Full HRTF & \(p<0.001 / <0.001\) & \(0.018\) & \([-0.039, 0.075]\) \\
\rowcolor{contribpurple}
Total angular & 6DoF & 20 & 2-Channel \(\leftrightarrow\) Full HRTF & \(p<0.001 / <0.001\) & \(0.057\) & \([0.009, 0.106]\) \\
\rowcolor{contribpurple}
Total angular & 6DoF & 20 & Azimuth-Only \(\leftrightarrow\) Full HRTF & \(p<0.001 / <0.001\) & \(0.004\) & \([-0.064, 0.072]\) \\
\rowcolor{contribpurple}
Total angular & 6DoF & 30 & 2-Channel \(\leftrightarrow\) Full HRTF & \(p<0.001 / <0.001\) & \(0.054\) & \([-0.039, 0.146]\) \\
\rowcolor{contribpurple}
Total angular & 6DoF & 30 & Azimuth-Only \(\leftrightarrow\) Full HRTF & \(p<0.001 / <0.001\) & \(-0.024\) & \([-0.097, 0.049]\) \\

Azimuth & 3DoF & 5 & 2-Channel \(\rightarrow\) Full HRTF & \(p=0.795\) & \(0.056\) & \([0.007, 0.104]\) \\
Azimuth & 3DoF & 5 & Azimuth-Only \(\rightarrow\) Full HRTF & \(p=0.011\) & \(-0.039\) & \([-0.091, 0.014]\) \\
Azimuth & 3DoF & 10 & 2-Channel \(\rightarrow\) Full HRTF & \(p=0.239\) & \(0.111\) & \([0.055, 0.167]\) \\
\rowcolor{contribpurple}
Azimuth & 3DoF & 10 & Azimuth-Only \(\leftrightarrow\) Full HRTF & \(p<0.001 / <0.001\) & \(-0.017\) & \([-0.068, 0.035]\) \\
\rowcolor{contribpurple}
Azimuth & 3DoF & 15 & 2-Channel \(\leftrightarrow\) Full HRTF & \(p=0.017 / <0.001\) & \(0.172\) & \([0.095, 0.249]\) \\
\rowcolor{contribpurple}
Azimuth & 3DoF & 15 & Azimuth-Only \(\leftrightarrow\) Full HRTF & \(p<0.001 / <0.001\) & \(0.003\) & \([-0.040, 0.045]\) \\
\rowcolor{contribpurple}
Azimuth & 3DoF & 20 & 2-Channel \(\leftrightarrow\) Full HRTF & \(p<0.001 / <0.001\) & \(0.150\) & \([0.063, 0.237]\) \\
\rowcolor{contribpurple}
Azimuth & 3DoF & 20 & Azimuth-Only \(\leftrightarrow\) Full HRTF & \(p<0.001 / <0.001\) & \(0.000\) & \([-0.057, 0.057]\) \\
\rowcolor{contribpurple}
Azimuth & 3DoF & 30 & 2-Channel \(\leftrightarrow\) Full HRTF & \(p<0.001 / <0.001\) & \(0.050\) & \([-0.027, 0.127]\) \\
\rowcolor{contribpurple}
Azimuth & 3DoF & 30 & Azimuth-Only \(\leftrightarrow\) Full HRTF & \(p<0.001 / <0.001\) & \(-0.028\) & \([-0.085, 0.029]\) \\

Azimuth & 6DoF & 5 & 2-Channel \(\rightarrow\) Full HRTF & \(p=0.402\) & \(0.053\) & \([-0.014, 0.120]\) \\
Azimuth & 6DoF & 5 & Azimuth-Only \(\rightarrow\) Full HRTF & \(p=0.012\) & \(-0.014\) & \([-0.073, 0.046]\) \\
Azimuth & 6DoF & 10 & 2-Channel \(\rightarrow\) Full HRTF & \(p=0.058\) & \(0.077\) & \([0.015, 0.138]\) \\
\rowcolor{contribpurple}
Azimuth & 6DoF & 10 & Azimuth-Only \(\leftrightarrow\) Full HRTF & \(p<0.001 / 0.007\) & \(-0.032\) & \([-0.106, 0.042]\) \\
\rowcolor{contribpurple}
Azimuth & 6DoF & 15 & 2-Channel \(\leftrightarrow\) Full HRTF & \(p=0.003 / <0.001\) & \(0.048\) & \([-0.028, 0.123]\) \\
\rowcolor{contribpurple}
Azimuth & 6DoF & 15 & Azimuth-Only \(\leftrightarrow\) Full HRTF & \(p<0.001 / <0.001\) & \(-0.050\) & \([-0.119, 0.019]\) \\
\rowcolor{contribpurple}
Azimuth & 6DoF & 20 & 2-Channel \(\leftrightarrow\) Full HRTF & \(p<0.001 / <0.001\) & \(0.043\) & \([-0.051, 0.136]\) \\
\rowcolor{contribpurple}
Azimuth & 6DoF & 20 & Azimuth-Only \(\leftrightarrow\) Full HRTF & \(p<0.001 / <0.001\) & \(-0.032\) & \([-0.105, 0.041]\) \\
\rowcolor{contribpurple}
Azimuth & 6DoF & 30 & 2-Channel \(\leftrightarrow\) Full HRTF & \(p<0.001 / <0.001\) & \(-0.001\) & \([-0.076, 0.074]\) \\
\rowcolor{contribpurple}
Azimuth & 6DoF & 30 & Azimuth-Only \(\leftrightarrow\) Full HRTF & \(p<0.001 / <0.001\) & \(-0.062\) & \([-0.120, -0.004]\) \\ 
\rowcolor{contribpurple}
Elevation & 3DoF & 5 & 2-Channel \(\leftrightarrow\) Full HRTF
& \(p<0.001 / <0.001\) & \(-0.017\) & \([-0.081, 0.048]\) \\
\rowcolor{contribpurple}
Elevation & 3DoF & 5 & Azimuth-Only \(\leftrightarrow\) Full HRTF
& \(p<0.001 / <0.001\) & \(0.033\) & \([-0.025, 0.091]\) \\
\rowcolor{contribpurple}
Elevation & 3DoF & 10 & 2-Channel \(\leftrightarrow\) Full HRTF
& \(p<0.001 / <0.001\) & \(0.008\) & \([-0.053, 0.070]\) \\
\rowcolor{contribpurple}
Elevation & 3DoF & 10 & Azimuth-Only \(\leftrightarrow\) Full HRTF
& \(p<0.001 / <0.001\) & \(0.036\) & \([-0.035, 0.107]\) \\
\rowcolor{contribpurple}
Elevation & 3DoF & 15 & 2-Channel \(\leftrightarrow\) Full HRTF
& \(p<0.001 / <0.001\) & \(0.008\) & \([-0.040, 0.057]\) \\
\rowcolor{contribpurple}
Elevation & 3DoF & 15 & Azimuth-Only \(\leftrightarrow\) Full HRTF
& \(p<0.001 / <0.001\) & \(0.014\) & \([-0.025, 0.052]\) \\
\rowcolor{contribpurple}
Elevation & 3DoF & 20 & 2-Channel \(\leftrightarrow\) Full HRTF
& \(p<0.001 / <0.001\) & \(0.011\) & \([-0.057, 0.079]\) \\
\rowcolor{contribpurple}
Elevation & 3DoF & 20 & Azimuth-Only \(\leftrightarrow\) Full HRTF
& \(p<0.001 / <0.001\) & \(0.011\) & \([-0.052, 0.074]\) \\
\rowcolor{contribpurple}
Elevation & 3DoF & 30 & 2-Channel \(\leftrightarrow\) Full HRTF
& \(p<0.001 / <0.001\) & \(-0.042\) & \([-0.085, 0.002]\) \\
\rowcolor{contribpurple}
Elevation & 3DoF & 30 & Azimuth-Only \(\leftrightarrow\) Full HRTF
& \(p<0.001 / <0.001\) & \(-0.028\) & \([-0.071, 0.015]\) \\

\rowcolor{contribpurple}
Elevation & 6DoF & 5 & 2-Channel \(\leftrightarrow\) Full HRTF
& \(p<0.001 / <0.001\) & \(-0.016\) & \([-0.088, 0.056]\) \\
\rowcolor{contribpurple}
Elevation & 6DoF & 5 & Azimuth-Only \(\leftrightarrow\) Full HRTF
& \(p<0.001 / <0.001\) & \(0.039\) & \([-0.032, 0.110]\) \\
\rowcolor{contribpurple}
Elevation & 6DoF & 10 & 2-Channel \(\leftrightarrow\) Full HRTF
& \(p<0.001 / <0.001\) & \(0.010\) & \([-0.050, 0.071]\) \\
\rowcolor{contribpurple}
Elevation & 6DoF & 10 & Azimuth-Only \(\leftrightarrow\) Full HRTF
& \(p<0.001 / <0.001\) & \(0.069\) & \([-0.003, 0.140]\) \\
\rowcolor{contribpurple}
Elevation & 6DoF & 15 & 2-Channel \(\leftrightarrow\) Full HRTF
& \(p<0.001 / <0.001\) & \(0.033\) & \([-0.015, 0.081]\) \\
\rowcolor{contribpurple}
Elevation & 6DoF & 15 & Azimuth-Only \(\leftrightarrow\) Full HRTF
& \(p<0.001 / <0.001\) & \(0.061\) & \([-0.008, 0.129]\) \\
\rowcolor{contribpurple}
Elevation & 6DoF & 20 & 2-Channel \(\leftrightarrow\) Full HRTF
& \(p<0.001 / <0.001\) & \(0.034\) & \([-0.018, 0.086]\) \\
\rowcolor{contribpurple}
Elevation & 6DoF & 20 & Azimuth-Only \(\leftrightarrow\) Full HRTF
& \(p<0.001 / <0.001\) & \(0.039\) & \([-0.017, 0.095]\) \\
\rowcolor{contribpurple}
Elevation & 6DoF & 30 & 2-Channel \(\leftrightarrow\) Full HRTF
& \(p<0.001 / <0.001\) & \(0.027\) & \([-0.007, 0.062]\) \\
\rowcolor{contribpurple}
Elevation & 6DoF & 30 & Azimuth-Only \(\leftrightarrow\) Full HRTF
& \(p<0.001 / <0.001\) & \(0.025\) & \([-0.027, 0.076]\) \\
\bottomrule
\end{tabular}
\end{table}

\clearpage
\end{document}